\documentclass[12pt,a4paper]{article}
\usepackage{graphicx}
\usepackage{amssymb,amsfonts,amsmath}
\usepackage{color}
\usepackage{cite}
\usepackage{bm}
\begin{document}

\newcommand{\EQUATION}{Equation~}
\newcommand{\EQ}{Eq.~}
\newcommand{\EQS}{Eqs.~}
\newcommand{\FIG}{Fig.~}
\newcommand{\FIGS}{Figs.~}
\newcommand{\SEC}{Sec.~}
\newcommand{\SECS}{Secs.~}

\newcommand{\comr}[1]{\textcolor{red}{#1}}
\newcommand{\comb}[1]{\textcolor{blue}{#1}}
\newcommand{\hiroshi}[1]{\textcolor{green}{\bf [Hiroshi: #1]}}
\newcommand{\naoki}[1]{\textcolor{green}{\bf [Naoki: #1]}}
\newcommand{\yoji}[1]{\textcolor{green}{\bf [Yoji: #1]}}
\newcommand{\edit}[1]{\textcolor{red}{\bf #1}}

\title{Collective fluctuations in networks of noisy components}
\author{Naoki Masuda${}^{1,2}$, Yoji Kawamura${}^{3}$,
 and Hiroshi Kori${}^{4,2*}$\\
\ \\
\ \\
${}^{1}$ 
Graduate School of Information Science and Technology,\\
The University of Tokyo,\\
7-3-1 Hongo, Bunkyo, Tokyo 113-8656, Japan
\ \\
${}^2$
PRESTO, Japan Science and Technology Agency,\\
4-1-8 Honcho, Kawaguchi, Saitama 332-0012, Japan
\ \\
${}^3$ Institute for Research on Earth Evolution,\\
Japan Agency for Marine-Earth Science and Technology,\\
3173-25 Showa-machi, Kanazawa-ku, Yokohama, Kanagawa 236-0001, Japan
\\
${}^4$ Division of Advanced Sciences, Ochadai Academic Production,\\
Ochanomizu University,\\
2-1-1, Ohtsuka, Bunkyo-ku, Tokyo 112-8610, Japan\\ 
\ \\
$^*$ Author for correspondence (kori.hiroshi@ocha.ac.jp)}

\setlength{\baselineskip}{0.77cm}
\maketitle

\newpage

\begin{abstract}
\setlength{\baselineskip}{0.77cm} 
Collective dynamics result from interactions among noisy dynamical
components. Examples include heartbeats, circadian rhythms, and various
pattern formations. Because of noise in each component, collective
dynamics inevitably involve fluctuations, which may crucially affect
functioning of the system. However, the relation between the
fluctuations in isolated individual components and those in collective
dynamics is unclear. Here we study a linear dynamical system of
networked components subjected to independent Gaussian noise and
analytically show that the connectivity of networks determines the
intensity of fluctuations in the collective dynamics. Remarkably, in
general directed networks including scale-free networks, the
fluctuations decrease more slowly with the system size than the standard
law stated by the central limit theorem. They even remain finite for a large system size
when global directionality of the network exists.
Moreover, such nontrivial behavior appears even in undirected networks
when nonlinear dynamical systems are considered. We demonstrate it with a coupled
oscillator system. 
\end{abstract}


\newpage

\section{Introduction}\label{sec:introduction}

Understanding fluctuations in dynamically ordered states and physical objects,
which consist of networks of interacting components, is an important issue in many
disciplines ranging from biology to engineering.
When each constituent component
of a system
is noisy due to, {\em e.g.},
thermal fluctuations, it generally occurs that
the entire system collectively fluctuates in time.
Such collective fluctuations may be
advantageous or disadvantageous in functioning of the systems
depending on situations.
For example, reduction in noise is likely to improve
information processing in retinal neural networks
\cite{Lamb76,Smith95,Devries02,Bloomfield04}.
Precision of biological circadian clocks
\cite{Wilders93,Enright80,Garcia04pnas,Herzog04}
may be improved by reduction in 
collective fluctuations (\textit{i.e.}, fluctuations in collective
activities).
On the other hand, maintaining a certain amount of fluctuations
in an ordered state
is advantageous for stochastic resonance
\cite{Gammaitoni98} and Brownian motors \cite{Reimann02}.

Despite the relevance of collective fluctuations in 
a variety of systems, 
theoretical frameworks that formulate collective fluctuations are missing.
The central limit theorem states that,
if the dynamical order is simply the averaged activity of noisy
components, the standard deviation of the collective fluctuation would
decrease with the number $N$ of noisy components as $N^{-1/2}$.
However, scaling is unclear in systems of interacting 
components. Clarifying the property of collective fluctuations in
such systems will give us insights into the mechanisms and design
principles underlying the regulation of noise in, for example, living
organisms and chemical reactions, and also into possible controls of fluctuations in collective
dynamics.

In this study, we analyze an ensemble of components subjected to
independent Gaussian noise that interact on general networks, including
complex networks and regular lattices.
We first consider a linear dynamical system,
which can be regarded as linearization of various systems, such as networks of periodic or chaotic oscillators \cite{kuramoto84,
pikovsky01}, the overdamped limit of elastic networks \cite{togashi07}, 
a consensus problem treated in control theory \cite{olfatisaber07}.
We show that collective fluctuations are
determined by the connectivity of networks.
It turns out that the scaling $N^{-1/2}$ 
is the tight lower bound,
which is obtained for undirected networks. General
directed networks yield a slower or nonvanishing decay of collective
fluctuations with an increase in $N$. 
We then argue such
nontrivial behavior appears even in undirected networks when 
nonlinear systems are considered. In particular, we show that linearization of
coupled nonlinear oscillator systems on undirected media yields
linear dynamics on asymmetric networks, such that the slow decay
of the collective fluctuation is relevant.

\section{Model and analysis}

Consider a network of $N$ components obeying
\begin{equation}
 \dot x_i= \sum_{j=1}^N w_{ij}(x_j-x_i)+\sqrt{D_i} \xi_i(t), \quad (1\le i\le N),
\label{eq:linear}
\end{equation}
where $x_i$ is the state (or the position) of the $i$th component,
$\sqrt{D_i}$ is the intensity of noise, 
$\xi_i$ is the independent Gaussian (generally colored) noise, and
$w_{ij}$ is the
intensity of coupling
and can be also regarded as originating
from the Jacobian matrix of underlying nonlinear dynamical systems
such as coupled oscillator systems that we consider later.
We allow negative weights and asymmetric coupling;
$w_{ij}$ can be negative or different from $w_{ji}$.
\EQUATION\eqref{eq:linear} is a multivariate Ornstein-Uhlenbeck process
\cite{Riskenbook,Vankampenbook}.

For convenience, we represent \EQ\eqref{eq:linear} as 
\begin{equation}
\dot {\bm x}=-L\bm x + \bm p,
\end{equation}
where ${\bm x} \equiv (x_1\; \ldots \;
x_N)^{\top}$ ($\top$ denotes the transpose), $\bm p \equiv (\sqrt{D_1}
\xi_1\; \ldots \; \sqrt{D_N} \xi_N)^{\top}$, and
$L=(L_{ij})$ is the asymmetric Laplacian defined
by 
$L_{ij}=\delta_{ij}\sum_{i^{\prime}\neq i}w_{ii^{\prime}}
-(1-\delta_{ij})w_{ij}$
\cite{Arenas08,mkk09njp}.
$L$ always has a zero eigenvalue
with the right eigenvector $\bm u \equiv (1\; \ldots\; 1)^{\top}$,
\textit{i.e.}, $L\bm u=0$.
This eigenvector is associated with a global translational
shift in state $\bm x$ and corresponds to the fact that
such a shift keeps \EQ\eqref{eq:linear} invariant.
We assume the stability of the ordered state represented by $x_1=\ldots=x_N$
in the absence of the noise (\textit{i.e.}, $D_i=0$ for
all $i$); the system relaxes to the ordered state
from any initial condition. This is equivalent to
assuming that the real parts of all the eigenvalues of 
$L$ are positive
except for one zero eigenvalue, \textit{i.e.}, $0\equiv
\lambda_1 < {\rm Re} \lambda_2 \le \ldots \le {\rm Re} \lambda_N$. 
This is a nontrivial condition for general networks with negative
weights.  However, for networks with only non-negative weights,
\textit{i.e.}, $w_{ij}\ge 0$ ($1\le i, j\le N$), this property holds
true when the network is strongly connected
or all the nodes are reachable by a directed
path from a single node \cite{Ermentrout92,Agaev00,Arenas08}.

We are concerned with collective fluctuations in dynamics given by
\EQ\eqref{eq:linear}.  To quantify their intensity, we decompose
$\bm x$ as
\begin{equation}
 \bm x(t) = y(t)\bm u + \bm \rho(t).
\label{eq:mode decomposition}
\end{equation}
$y(t)$ describes the one-dimensional component along $\bm u$, and $\bm
\rho(t)$ is the ($N-1$)-dimensional remainder mode.
Note that $y(t) =
\bm v \bm x(t)$, where the row vector $\bm v\equiv (v_1 \ldots v_N)$ is
the left eigenvector of $L$ corresponding to the zero
eigenvalue, \textit{i.e.}, $\bm v L=0$,
and is normalized as
$\bm v \bm u =1$, \textit{i.e.}, $\sum_{i=1}^N v_i=1$
(see Appendix~A for detailed descriptions).
We call $y(t)\bm u$ the collective mode.
In the absence of noise, the dynamical equation for $y(t)$ is given by
\begin{equation}
\dot y = \bm v \dot {\bm x} = -
\bm v L \bm x =0.
\end{equation}
Therefore, $y(t)$ is a
conserved quantity of the dynamics.
The remainder mode $\bm \rho(t)$ is associated with relative
motions among the components. Because of the stability assumption,
$\bm \rho(t)$ asymptotically vanishes with characteristic time
$({\rm Re}~\lambda_2)^{-1}$. Therefore, all the values of
$x_i$ ($1\le i \le N$) eventually go to the same value $y$
that is determined by the initial condition, \textit{i.e.}, $y=\bm v
\bm x(0)$.

In the presence of noise,
we obtain
\begin{equation}
\dot y = \bm v \dot {\bm x} = \bm v (-L \bm x + \bm
p) = \bm v \bm p = \sum_{i=1}^N v_i \sqrt{D_i}
\xi_i(t).
\end{equation}
Because $\xi_i$ is the independent Gaussian noise, this
equation reduces to
\begin{equation}
\dot y(t) =  \sqrt{\sum_{i=1}^N  v_i^2 D_i} \xi (t) \equiv \sigma \xi (t),
\label{eq:dynamics of collective mode}
\end{equation}
where $\xi(t)$ is the Gaussian noise having the same statistical
property as that of each $\xi_i(t)$.  Thus, $y(t)$ performs the Brownian
motion with effective noise strength $\sigma$ and is unbounded. 
The remainder mode $\bm \rho (t)$ fluctuates around zero
because of
its decaying nature. 
Therefore, the
long-time behavior of $x_i(t) = y(t) + \rho_i(t)$
is approximately described by a single variable $y(t)$ for any $i$.
We denote as $\sigma$, which depends on the structure of the network,
the intensity of collective fluctuations.
$\sigma$ can be calculated for given network.

In practice, the average activity of the population, $\bar x\equiv \sum_{i=1}^N
x_i/N$, but not the activity at individual nodes, may be observed.
Because $\bar x = y + \sum_{i=1}^N \rho_i/N$ and 
$\sum_{i=1}^N \rho_i/N$ can be
neglected in a long run,
$\sigma$ also characterizes
the fluctuations of $\bar x$.

\section{Collective fluctuations in various networks}

\subsection{General properties}

We assume for simplicity that $D_i=1$ ($1\le i\le N$) so
that $\sigma=\sqrt{\sum_{i=1}^N v_i^2}$. 
It is straightforward to extend the following results to the case of
heterogeneous $D_i$.
The vector $\bm v$ is uniform, \textit{i.e.},
$v_i=1/N$ ($1\le i\le N$) if and only if $k_i^{\rm in}=k_i^{\rm out}$
($1\le i\le N$), where $k_i^{\rm in}\equiv \sum_{j=1}^N w_{ij}$ and $k_i^{\rm
out}\equiv \sum_{j=1}^N w_{ji}$ are indegree and outdegree, respectively 
\cite{mkk09njp}.
Undirected networks satisfy this condition. In this case,
we obtain $\sigma=N^{-1/2}$, which agrees with the central limit
theorem. The normalization condition
$\sum_{i=1}^N v_i=1$ guarantees that $\sigma\ge
N^{-1/2}$ for any $\bm v$. Therefore, undirected
networks are the best for reducing collective fluctuations. In the case of
directed or asymmetrically weighted networks, $v_i$ is generally
heterogeneous, and $\sigma > N^{-1/2}$.
We will show later
that this is also the case for nonlinear 
systems on undirected networks.
When the weight $w_{ij}$ is nonnegative
for any $i$ and $j$, the Perron--Frobenius theorem guarantees that $v_i$
is nonnegative for all $i$ \cite{Hornbook}.  In this case, we obtain
\begin{equation}
\frac{1}{\sqrt{N}}\le \sigma \le 1.
\end{equation}
The case $\sigma=1$
is realized by a feedforward network,
in which a certain component $i_0$ has no inward connection
(\textit{i.e.}, $k_{i_0}^{\rm in}=0$).  Then, $v_{i_0}=1$ and $v_i= 0$
for $i \neq i_0$, which yields $\sigma=1$ irrespective of $N$; the
collective fluctuations are not reduced at all with an increase in $N$. 
When negative weights are allowed, some elements of $\bm v$ may
assume negative values. Then, $\sigma$ may be larger than
$1$, in which case
collective fluctuations are larger than individual noise.

We note that  $\sigma^{2}$ is the so-called inverse
participation ratio \cite{Derrida87}.
$\sigma^{-2}$ can be interpreted as the effective
number of components that participate in collective
activities; the remaining components are slaved.

\subsection{Directed scale-free networks}
\begin{figure}
\begin{center}
\includegraphics[width=8cm]{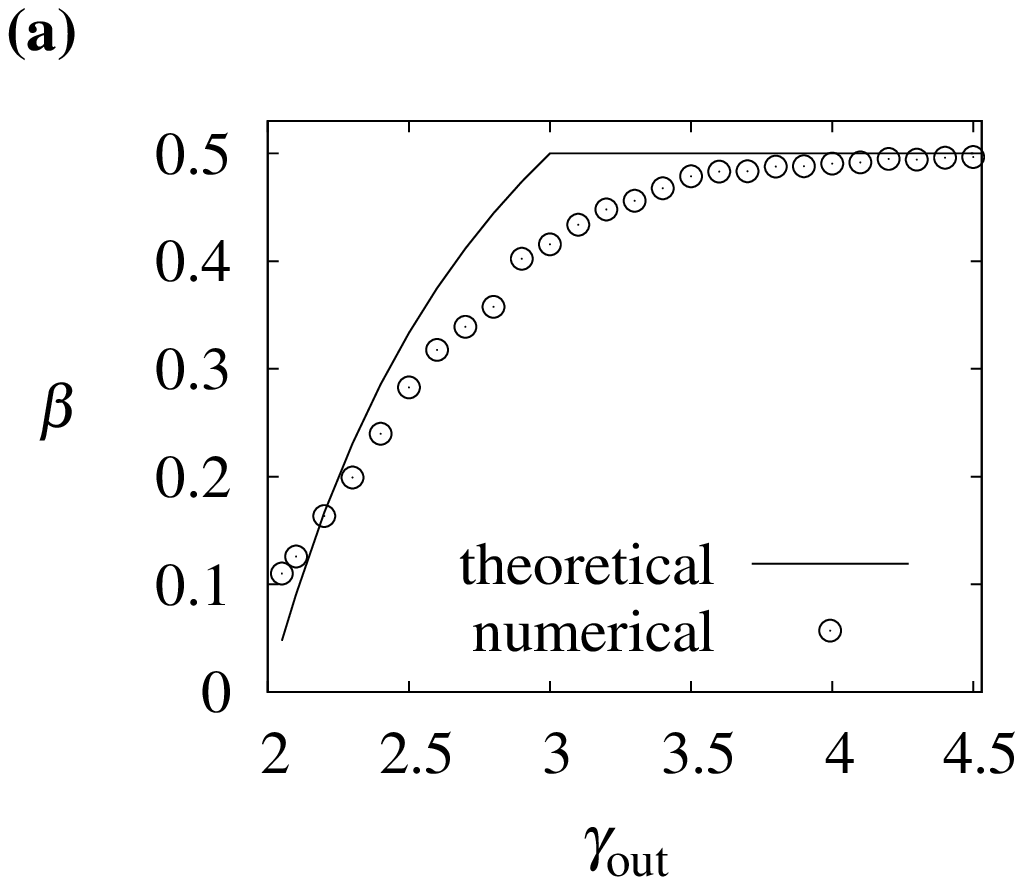}
\includegraphics[width=8cm]{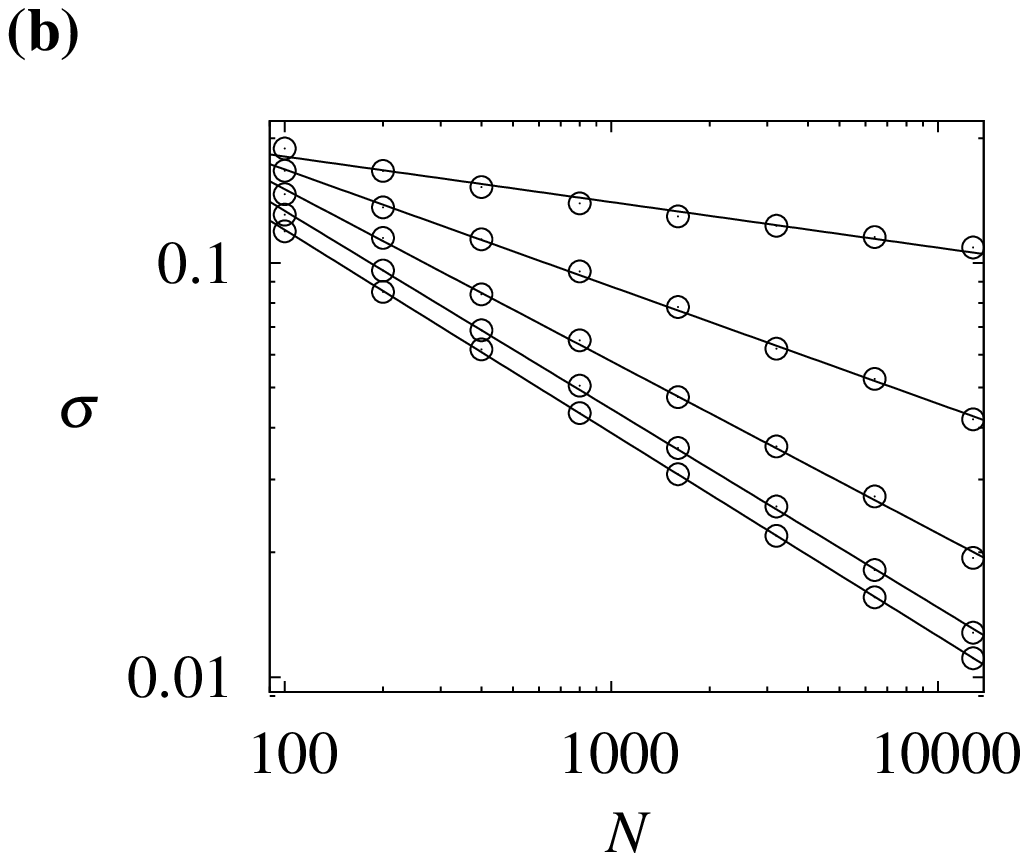}
\caption{(a) Scaling exponent $\beta$ for $\sigma\propto N^{-\beta}$ in
scale-free networks with $\gamma_{\rm in}=\gamma_{\rm out}$.
The solid line is the theoretical prediction
given by
\EQ\eqref{eq:sf theory}.  The open circles are obtained
numerically as follows. For
each network, we calculate
the eigenvector $\bm v$ to obtain $\sigma$.  Then, $\beta$
is determined from the best linear fit between $\beta$ and $N$ on the
logarithmic scale, as described in (b). (b) Determination of 
$\beta$.
Data points are generated as 
an average value of $\beta$ for each of $N=100$,
200, 400, $\ldots$, 12800.
The results obtained from the direct
numerical simulations are shown by circles.
For demonstration, the results for
$\gamma_{\rm in}=\gamma_{\rm out}=2.05$, 2.5, 3, 3.5, and 4 are shown.
By assuming $\sigma\propto N^{-\beta}$,
we regress $\log\sigma$ against $\log N$
by the best linear fit (solid lines). The slope gives an estimate
of $-\beta$. 
The Pearson correlation coefficient is
large ($>$ 0.99) for each value of $\gamma_{\rm out}$
analyzed in (a).
}
\label{fig:sigma sf}
\end{center}
\end{figure}

We demonstrate our theory by using some example networks.
First, we consider directed scale-free networks,
schematically shown in \FIG\ref{fig:schem nets}(a)
in which
$k_i^{\rm in}$ and $k_i^{\rm out}$ independently
follow the distributions
$p(k^{\rm in})\propto k^{-\gamma_{\rm in}}$
and $p(k^{\rm out})\propto k^{-\gamma_{\rm out}}$, respectively.
By assuming that the values of $v_i$ of adjacent nodes are independent
of each other, we obtain
\begin{equation}
\sum_{j=1}^N w_{ji} v_j \approx
\sum_{j=1}^N w_{ji} \bar{v} = k_i^{\rm out} \bar{v},
\end{equation}
where
\begin{equation}
\bar{v}\equiv \frac{\sum^N_{i=1}v_i}{N}=\frac{1}{N}.
\end{equation}
Therefore,
\begin{equation}
v_i
=\frac{\sum_{j=1}^N w_{ji} v_j}{\sum_{j=1}^N w_{ij}} \approx 
\frac{k_i^{\rm out}/k_i^{\rm in}}{\sum_{j=1}^N\left(k_j^{\rm out}/k_j^{\rm
in}\right)}.
\end{equation} 
This approximation is sufficiently accurate
for uncorrelated networks \cite{mkk09njp}.
For $p(k^{\rm in})\propto k^{-\gamma_{\rm in}}$
and $p(k^{\rm out})\propto k^{-\gamma_{\rm out}}$,
we obtain
\begin{equation}
v_i\approx \frac{k_i^{\rm out}/k_i^{\rm
in}}{N\left<k^{\rm out}\right>\left<(k^{\rm in})^{-1}\right>}
\end{equation}
and
\begin{equation}
\sigma\approx
\sqrt{\frac{\left<\left(k^{\rm out}\right)^2\right>
\left<\left(k^{\rm in}\right)^{-2}\right>}
{N\left<k^{\rm out}\right>^2\left<\left(k^{\rm in}\right)
^{-1}\right>^2}},
\end{equation}
where $\left< \cdot \right>$ is
the ensemble average.  
When $\gamma_{\rm out}<2$, a winner-take-all
network is generated \cite{Albert02rmp,Dorogovtsev08rmp}, and there exists
a node $i$
such that $k_i^{\rm out}=O(N)$ and $v_i=O(1)$.  When
$\gamma_{\rm out}\ge 2$, the extremal criterion results in
the maximum degree increasing with $N$
as $N^{1/(\gamma_{\rm out}-1)}$ ($\gamma_{\rm out}\ge 2$) in many networks
\cite{Newman05comp,Dorogovtsev08rmp}. Then,
we obtain \cite{Sood08pre}
\begin{equation}
\left<k^{\rm out}\right>\propto
\left\{\begin{array}{ll}
N^{2-\gamma_{\rm out}},& (\gamma_{\rm out}< 2),\\
\ln N,& (\gamma_{\rm out}=2),\\
O(1),& (\gamma_{\rm out}>2),
\end{array}\right.
\end{equation}
\begin{equation}
\left<\left(k^{\rm out}\right)^2\right>\propto
\left\{\begin{array}{ll}
N^{-\gamma_{\rm out}+3},& (\gamma_{\rm out}< 2),\\
N^{(-\gamma_{\rm out}+3)/(\gamma_{\rm out}-1)},& (2\le 
\gamma_{\rm out}< 3),\\
\ln N,& (\gamma_{\rm out}=3),\\
O(1),& (\gamma_{\rm out}>3),
\end{array}\right.
\end{equation}
%
%
and
\begin{equation}
\left<\left(k^{\rm in}\right)^{-1}\right>,
\left<\left(k^{\rm in}\right)^{-2}\right>
=O(1).
\end{equation}
Therefore, we obtain
\begin{equation}
\sigma\propto \left\{\begin{array}{ll}
1,& (\gamma_{\rm out}<2),\\
1/\ln N,& (\gamma_{\rm out}=2),\\
N^{-1+(\gamma_{\rm out}-1)^{-1}},& (2\le \gamma_{\rm out}<3),\\
N^{-1/2}(\ln N)^{1/2},& (\gamma_{\rm out}= 3),\\
N^{-1/2},& (\gamma_{\rm out}> 3).
\end{array}\right. 
\label{eq:sf theory}
\end{equation}
The fairly heterogeneous case
$\gamma_{\rm out}<2$, in which the average outdegree diverges as
$N\to\infty$, effectively yields a feedforward network.  The case
$\gamma_{\rm out}\ge 3$, where the second moment of the outdegree
converges for $N\to\infty$, reproduces the central limit theorem.  The
latter result is shared by the directed version of the conventional
random graph.  The case $2\le\gamma_{\rm out}<3$ yields a nontrivial
dependence of $\sigma$ on $N$.
In \FIG\ref{fig:sigma sf}(a), we compare the scaling exponent
$\beta$, where $\sigma\propto N^{-\beta}$ obtained from
the theory (solid line; \EQ\eqref{eq:sf theory}) and
numerical simulations of
the configuration model \cite{Boccaletti06,Albert02rmp} 
with the power-law degree distribution with minimum degree 3 (open circles). 
The fitting procedure is explained in
 \FIG\ref{fig:sigma sf}(b).
Equation~\eqref{eq:sf theory} roughly explains numerically obtained values of
$\beta$.

\subsection{Directed lattices}
\begin{figure}
\begin{center}
\includegraphics[width=8cm]{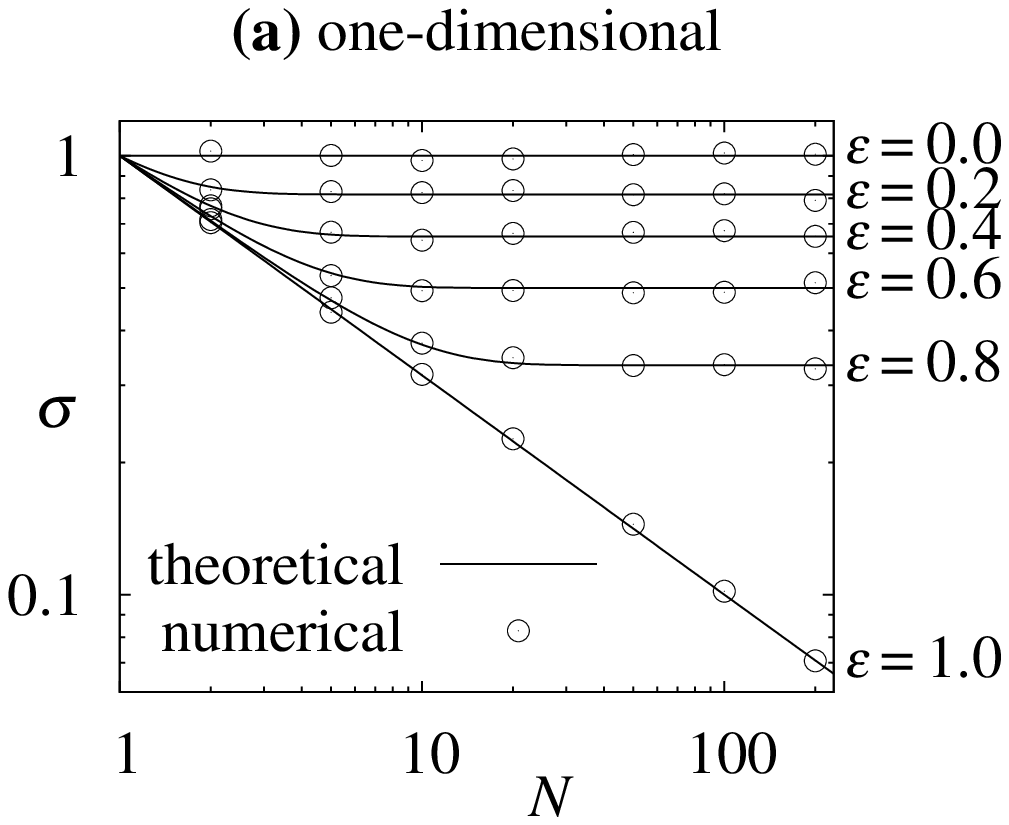}
\includegraphics[width=8cm]{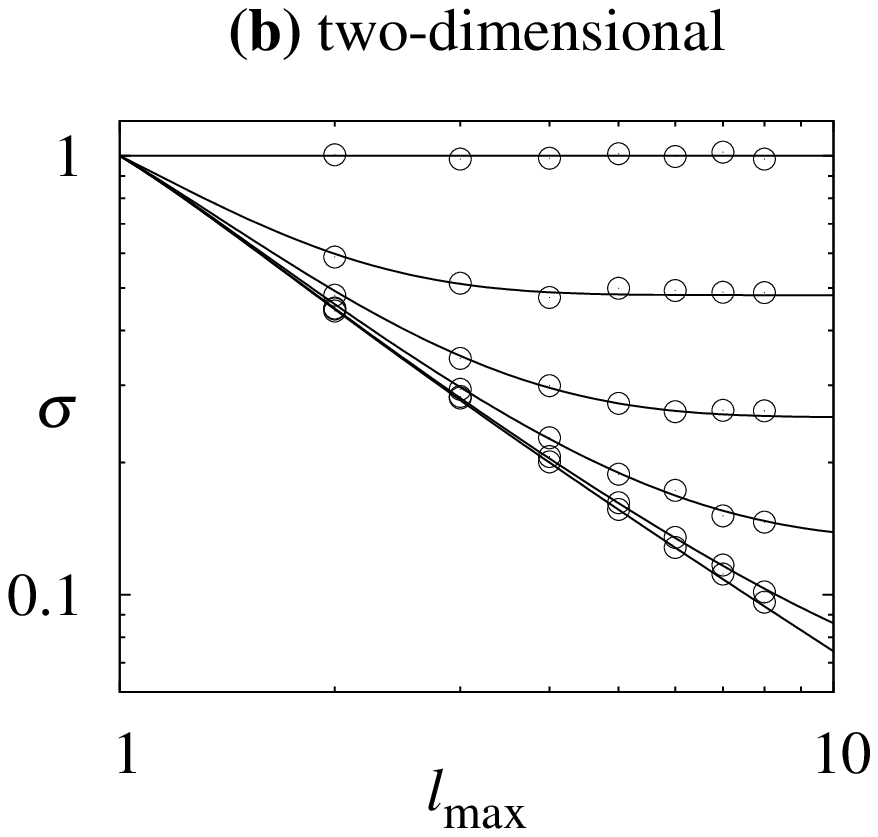}
\caption{Collective fluctuations for
(a) the directed one-dimensional chain and
(b) the directed two-dimensional lattice for various $N$ and $\epsilon$. 
The solid lines and the circles represent the theoretical and
the numerical results, respectively.
In (b), $\ell_{\max}$
is the maximum distance from the center of the lattice.
For both networks, we set $D_i=1$ ($1\le i\le N$)
and simulate \EQ\eqref{eq:linear} 
with the initial condition
$x_i(t=0)=0$ ($1\le i\le N$). We measured $\sigma$
as the standard deviation of
$\bar{x}(t=10200)-\bar{x}(t=200)$
obtained by conducting 2000 trials, which is then normalized
by $\sqrt{10000}$.
We disregard the first 200 time units as transient.}
\label{fig:sigma linear}
\end{center}
\end{figure}

The second example is the directed one-dimensional chain of $N$ nodes
depicted in \FIG\ref{fig:schem nets}(b).
We set $w_{i+1,i}=1$ ($1\le
i\le N-1$), $w_{i-1,i}=\epsilon$ ($2\le i\le N$), and $w_{j,i}=0$
($j\neq i-1, i+1$). For this network, by solving $\bm v L=0$, we analytically obtain
\begin{equation}
v_i = \frac{(1 -
\epsilon)\epsilon^{i-1}}{1 - \epsilon^N},\quad (1\le i\le N)
\end{equation}
and
\begin{equation}
  \sigma = \sqrt{\frac{1 - \epsilon}{1 + \epsilon} \, \frac{1 + \epsilon^N}{1 - \epsilon^N}}.
\end{equation}
Values of $\sigma$ for various $\epsilon$ and $N$ are plotted by
solid lines in \FIG\ref{fig:sigma linear}(a). Interestingly, for
$\epsilon\neq 1$, $\lim_{N \to \infty} \sigma = \sqrt{(1 - \epsilon)/(1
+ \epsilon)}$; $\sigma$ is nonvanishing.
  We have also analytically derived $\sigma$ for
directed $d$-dimensional lattices (see Appendix~B).  The
results for the two-dimensional lattice
depicted in \FIG\ref{fig:schem nets}(c)
are plotted by solid lines in \FIG\ref{fig:sigma
linear}(b). To confirm our theory, we also carried out direct
numerical simulations of \EQ\eqref{eq:linear} with Gaussian white noise
for these directed lattices. The results
indicated by circles in \FIG\ref{fig:sigma
linear} indicate an excellent agreement with our theory. 

A similar result is obtained for the Cayley tree (see Appendix~C).

\section{Oscillator dynamics}

\begin{figure}
\begin{center}
\includegraphics[width=8cm]{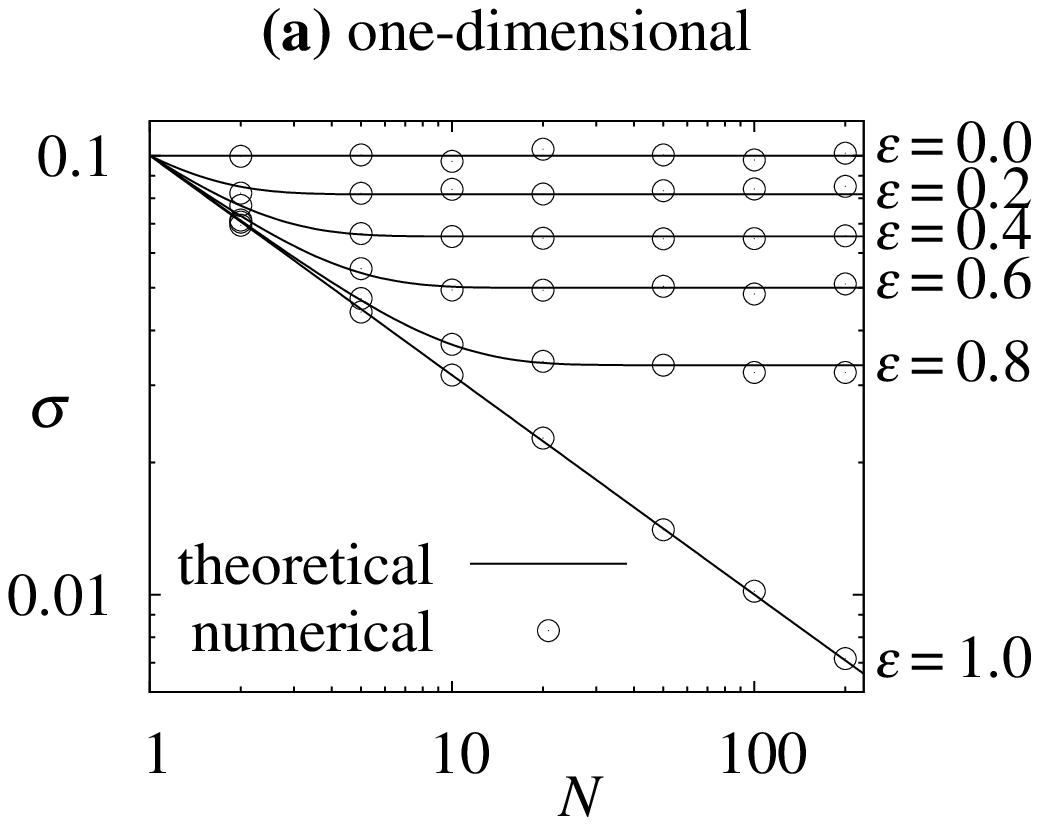}
\includegraphics[width=8cm]{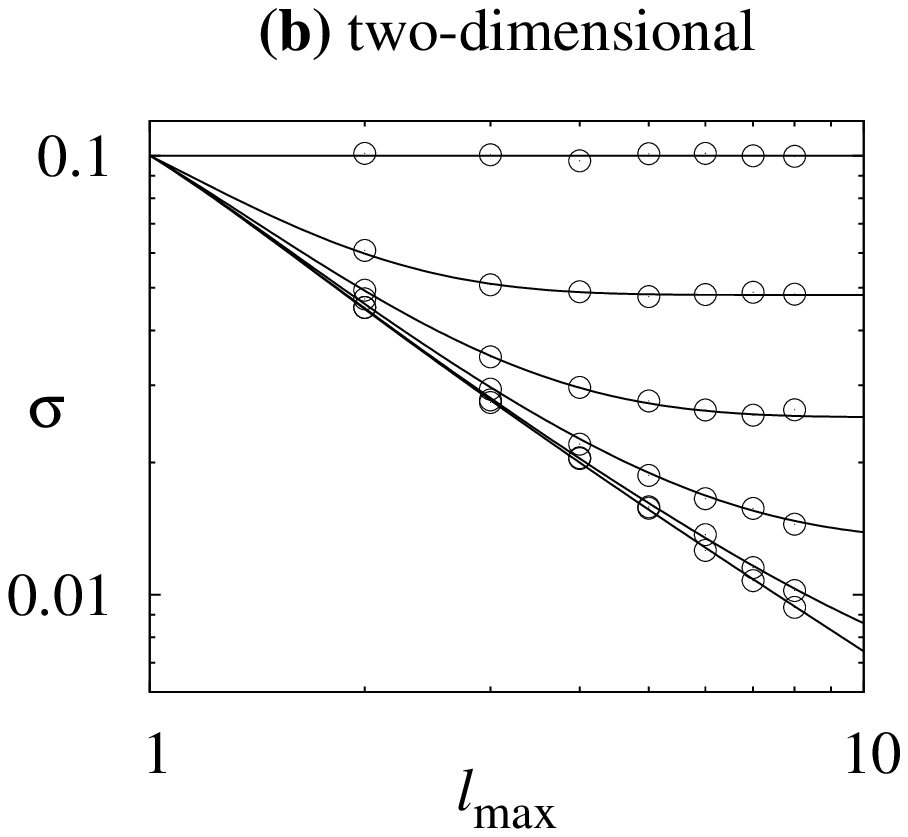}
\caption{Collective fluctuation for coupled phase oscillators in (a)
directed one-dimensional chain and (b) directed
two-dimensional lattice. The solid lines
represent the theoretical results, and the circles represent the numerical results
obtained by the direct numerical simulations of \EQ\eqref{eq:phase}
with $f(\phi)=\sin \phi$.
We set $D_i=0.01$ ($1\le i\le N$) and 
start with $x_i=0$ ($1\le i\le N$). We
measure $\sigma$
as the standard deviation of
$\bar{x}(t=10200) - \bar{x}(t=200)$
obtained by conducting 2000 trials, which is then normalized by
$\sqrt{10000}$.}
\label{fig:sigma osc}
\end{center}
\end{figure}

As an application
of our theory to nonlinear systems, we examine noisy
and rhythmic components.
As a general,
tractable, yet realistic model, we consider a network of phase
oscillators \cite{winfree67, kuramoto84, kiss07}, whose dynamical
equation is given by
\begin{equation}
 \dot \phi_i = \omega_i +
\sum_{j=1}^N A_{ij} f(\phi_j-\phi_i)+ \sqrt{D_i} \xi_i(t),
\quad (1\le i\le N),
 \label{eq:phase}
\end{equation}
where $\phi_i\in [0,2\pi)$ and $\omega_i$ are the phase and the
intrinsic frequency of the $i$th oscillator, respectively, $A_{ij}$ is
the intensity of coupling, and $f(\cdot)$
is a $2\pi$--periodic function. We assume
that, in the absence of noise, all the oscillators are in a
fully phase-locked state,
\textit{i.e.},
$\phi_i(t) = \Omega t + \psi_i$, where $\Omega$ and $\psi_i$ are
the constants derived from $\dot \phi_i = \Omega$ ($1\le i \le N$).
Under sufficiently weak noise,
we can linearize \EQ\eqref{eq:phase} around
the phase-locked state. Letting $x_i=\phi_i-(\psi_i+\Omega t)$, we
obtain \EQ\eqref{eq:linear}, where $w_{ij} = A_{ij}
f^{\prime}(\psi_j-\psi_i)$ is the effective weight.
The validity of linearizing \EQ\eqref{eq:phase} for small noise intensity
is tested by carrying out direct numerical
simulations of \EQ\eqref{eq:phase} 
with $\omega_i=\omega$ ($1\le i\le N$) and  $f(\phi)=\sin \phi$.
The relationship $\sigma\approx \sqrt{\sum_{i=1}^N
D_i v_i^2}$ is satisfied in the directed
one- and two-dimensional lattices,
as shown in \FIG\ref{fig:sigma osc}(a) and (b), respectively.

\begin{figure}
\begin{center}
\includegraphics[width=8cm]{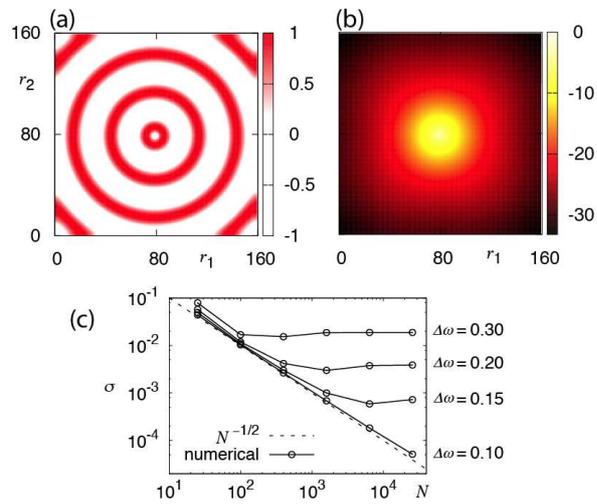}
\caption{(color online) Numerical results of the coupled oscillators on the
two-dimensional undirected lattice.
(a) Snapshot of $\sin \phi_i$ and (b) eigenvector $\bm v$ (log scale)
for $\Delta\omega=0.3$, where $r_1$
 and $r_2$ denote the spatial coordinates. (c) The dependence of
 $\sigma$ on system size $N$.
}
\label{fig:phase eigenvector}
\end{center}
\end{figure}

When there is some dispersion in $\psi_i$ in a phase-locked state, the relation
$\sigma\approx N^{-1/2}$ may be violated even in
undirected networks.
This is because the effective weight is
generally asymmetric (\textit{i.e.}, $w_{ij}\neq w_{ji}$)
unless $f(\cdot)$ is an exact odd function.
In reality, $f(\cdot)$ is usually not an odd
function \cite{kuramoto84,brown04,galan05,kiss07}.
As an example, we consider target
patterns (\textit{i.e.}, concentric traveling waves), which naturally
appear in spatially extended oscillator systems \cite{kuramoto84,
mikhailov06}.
We carry out direct numerical
simulations of \EQ\eqref{eq:phase} on 
the two-dimensional  {\em undirected} lattice with linear length $\sqrt{N}=50$,
$f(\phi)=\sin(\phi-\alpha)+\sin \alpha$, and $\alpha=\pi/4$.
Such a function may be
analytically derived from a general class of coupled oscillators
\cite{kuramoto84}, and it approximates a variety of real systems
\cite{galan05, tsubo07, kiss07}. 
We set $\omega_i=\omega_0 +
\Delta\omega$ $(\Delta\omega \ge 0)$ for $4 \times 4$ pacemaker
oscillators in the center and $\omega_i=\omega_0$ for the other
oscillators, where $\omega_0$ is arbitrary and set to $1$.
A target pattern is formed when there is sufficient heterogeneity in the
intrinsic frequency
\cite{kuramoto84}. 
A region with high intrinsic frequency acts as a pacemaker.
A snapshot for $\Delta \omega=0.3$ is shown in
\FIG\ref{fig:phase eigenvector}(a).
As observed, the radial phase gradient is
approximately constant, which makes the effective network
similar to the directed two-dimensional lattice
depicted in \FIG\ref{fig:schem nets}(c).
Therefore, as shown in \FIG\ref{fig:phase eigenvector}(b), $v_i$ calculated
numerically decreases almost
exponentially with the distance from the center.
We find that the dependence of $\sigma$ on $N$, shown in 
\FIG\ref{fig:phase eigenvector}(c), 
is similar to that for directed lattices. 

We emphasize that the network is
undirected (\textit{i.e.}, $A_{ij}=A_{ji}$). We have also theoretically
confirmed that our results are valid for the continuous oscillatory
media under spatial block noise, which models chemical
reaction--diffusion systems (see Appendix~D).

\section{Conclusions}

In summary, we
have obtained the analytical relationship between collective
fluctuations and the structure of networks. In undirected networks, the
fluctuations decrease with the system size $N$ as $N^{-1/2}$; this
result agrees with the central limit theorem. In general directed
networks, the collective fluctuations decay more slowly. For example, in
directed scale-free networks, we obtain $N^{-\beta}$ with $0 < \beta <
1/2$.
In networks with global directionality, the fluctuations do not
vanish for a large system size.
We have also demonstrated that such nontrivial
dependence appears even in undirected networks when nonlinear
systems are considered.
We have focused on systems of nonleaky components.
Results for coupled leaky components will be reported elsewhere.

Our results are distinct from earlier results demonstrating the breach
of the central limit theorem due to heavy-tailed noise
\cite{Bouchaud90} or the correlation
between the noise in different elements \cite{Kaneko90,Zohary94}.

Finally,
because our theory is based on a general
linear model,
it can be tested in a variety of experimental
systems.
An ideal experimental protocol is provided by
photo-sensitive Belousov-Zhabotinsky reaction systems, in which the
heterogeneity, noise intensity, and system size can be precisely
controlled by light stimuli \cite{mikhailov06}. 
Experiments with coupled oscillatory cells, such as cardiac cells and neurons under
an appropriate condition, would be also interesting.

\section*{Acknowledgments}

We thank Istvan Z. Kiss, Norio Konno, Yoshiki Kuramoto and Ralf
T\"{o}njes for their valuable discussions.
%
N.M. acknowledges
the support through the Grants-in-Aid for Scientific Research
(Nos. 20760258 and 20540382) from MEXT, Japan.


\renewcommand{\theequation}{S.\arabic{equation}}

\section*{Appendix A: Derivation of the collective mode}

To derive the collective mode $y(t)\bm u$,
we note that there exists a nonsingular
matrix $P$ such that 
$\tilde{L} \equiv P^{-1}LP$ is its Jordan
canonical form \cite{Hornbook,Arenas08}.
We assume that $\tilde{L}_{11}=\lambda_1=0$ and
$\tilde{L}_{1i}= \tilde{L}_{i1}=0$ ($2\le i\le N$) without loss of
generality.  The submatrix $(\tilde{L}_{ij})$ ($2\le i,j\le N$)
corresponds to the $N-1$ modes with the eigenvalues
$\lambda_2,\ldots, \lambda_N$. 
Because the first column of $LP=P\tilde{L}$ is equal to
$(0\;\ldots\; 0)^{\top}$,
the first column of $P$ is equal
to the right eigenvector of $L$ corresponding to
$\lambda_1=0$, \textit{i.e.},
$\bm u = (1\;\ldots\; 1)^{\top}$. Because the first row of
$P^{-1}L=\tilde{L}P^{-1}$ is equal to $(0\;\ldots\; 0)$,
the first row of $P^{-1}$ is equal to
the left eigenvector of $L$ corresponding to $\lambda_1=0$,
\textit{i.e.}, $\bm v = (v_1\; v_2\;\ldots\; v_N)$.  The normalization is
given by $\sum_{i=1}^N v_i=1$.
Under the variable change $(y\; \bm y_{\rm r})
\equiv P^{-1}\bm x
\in \mathbb{R}^N$, where $y\in \mathbb{R}$ and $\bm y_{\rm r}\in
\mathbb{R}^{N-1}$,
the coupling term is transformed into
$-\tilde{L} (y\; \bm y_{\rm r})$. Then, 
in the absence of the dynamical noise,
$y=\sum_{i=1}^N v_i x_i$ is a conserved quantity, which is the collective mode.
$\bm \rho(t)$ in
\EQ\eqref{eq:mode decomposition}
is given by $P_{\rm r}\bm y_{\rm r}$, where 
$P_{\rm r}$ is the $N$ by $N-1$ matrix satisfying
$P=(\bm u\; P_{\rm r})$.

\section*{Appendix B:
Collective fluctuations in regular lattices with arbitrary dimensions}

\begin{figure}
\begin{center}
\includegraphics[width=15cm]{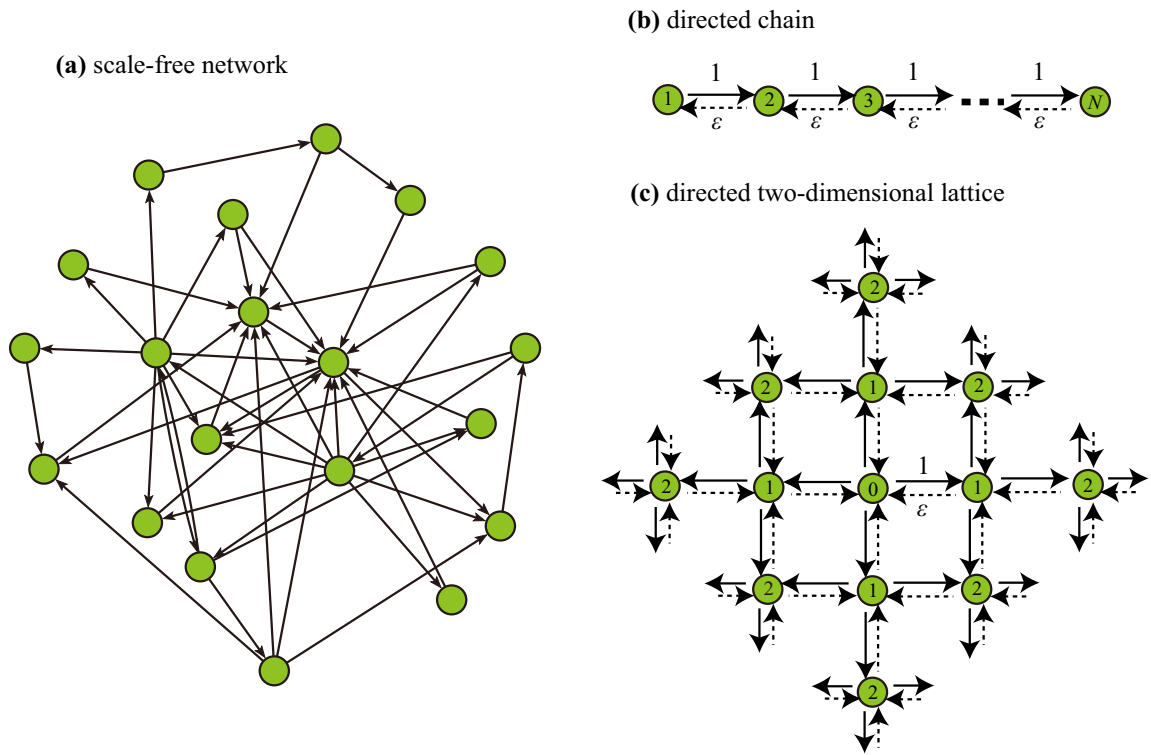}
\caption{Schematic of (a) directed scale-free network,
(b) directed chain, and (c) directed two-dimensional 
lattice. The numbers in (b) indicate the indices of the nodes, while
those in (c) indicate the layer index.}
\label{fig:schem nets}
\end{center}
\end{figure}

Consider a directed 
two-dimensional square lattice with a root node. As depicted in
\FIG\ref{fig:schem nets}(c), the edges descending from the
root node and those approaching the root node
in terms of the graph-theoretic distance are given weight 1
and $\epsilon$ ($0\le\epsilon\le 1$), respectively.
We define layers such that the layer $\ell$ ($\le \ell_{\max}$)
is occupied by the
nodes whose distance from the root node is equal to $\ell$. 
Layer 0
contains the root node only, and layer $\ell$ ($\ge 1$) contains
$4\ell$ nodes. We consider the lattice within a finite range
specified by $\ell\le \ell_{\max}$.
Note the difference from the case of the one-dimensional chain examined in
the main text (\FIG\ref{fig:schem nets}(b)),
where the root node is located
at the periphery of the chain.  However, the scaling of
$\sigma$ is not essentially
affected by this difference.

The symmetry guarantees that the four nodes in layer 1 have the same
value of $v_i$.  
Consider a node in \FIG \ref{fig:schem nets}(c)
that is labeled 2 and adjacent to two nodes labeled 1. There are
four such nodes.
The equation in $\bm v
L=0$ corresponding to this node is given by
 $(2+2\epsilon)v_2=2\epsilon v_1+2v_3$.
The other four nodes labeled 2 in \FIG \ref{fig:schem nets}(c)
yield a different equation
$(1+3\epsilon)v_2=\epsilon v_1+3v_3$.
Similarly, we obtain $(2+2\epsilon)v_\ell=2\epsilon v_{\ell-1}+
2v_{\ell+1}$ for all but four nodes in layer $\ell$. 
The other four nodes satisfy
 $(1+3\epsilon)v_\ell=\epsilon v_{\ell-1}+
3v_{\ell+1}$. Despite this inhomogeneity,
$v_{\ell}\propto \epsilon^{\ell}$ satisfies all these equations.
By counting the number of nodes in each layer,
the properly normalized solution is given by
\begin{equation}
  v_\ell = \left[ T_{\ell_{\max}}^{(2)}(\epsilon) \right]^{-1} \, \epsilon^\ell, \qquad
  (0\le \ell\le \ell_{\max}),
\end{equation}
and
\begin{equation}
  \sigma
  = \frac{\sqrt{T_{\ell_{\max}}^{(2)}(\epsilon^2)}}
{T_{\ell_{\max}}^{(2)}(\epsilon)},
\label{eq:sigma z2}
\end{equation}
where 
\begin{eqnarray}
  T_{\ell_{\max}}^{(2)}(z) &=& 1+ 4\sum_{\ell=1}^{\ell_{\max}} \ell z^\ell
\nonumber\\
&=& \frac{(1+z)^2 - 4[ 1 + {\ell_{\max}} (1 - z) ] z^{\ell_{\max}+1}}
{(1 - z)^2}.
\end{eqnarray}
The difference between the one- and two-dimensional cases
lies in the number of nodes in each layer, which affects the normalization
of $v_\ell$ and hence the value of $\sigma$.
In the limit of a purely feedforward network,
$\sigma$ is independent of the system size,
\textit{i.e.}, $\lim_{\epsilon \to 0} \sigma = 1$.
In the case of undirected networks, the
central limit theorem is recovered,
\textit{i.e.}, $\lim_{\epsilon \to 1} \sigma = N^{-1/2}$.
In the limit of infinite space,
we obtain 
\begin{equation}
\lim_{{\ell_{\max}}\to\infty}\sigma=
\frac{(1-\epsilon)(1+\epsilon^2)}{(1+\epsilon)^3}.
\end{equation}

For a general dimension $d$,
layer 0 has a single root node, and layer $\ell$
($1\le \ell\le \ell_{\max}$) has 
\begin{equation*}
N_{\ell}^{(d)} \equiv 
\sum_{d^{\prime}=1}^{\ell} \frac{d!}{d^{\prime}!(d-d^{\prime})!}
\frac{(\ell-1)!}{(d^{\prime}-1)!(\ell-d^{\prime})!}
2^{d^{\prime}}
\end{equation*}
nodes. $d^{\prime}$ is the number of coordinates
among the $d$ coordinates to which nonzero values are assigned, and
the factor $2^{d^{\prime}}$ takes care of the fact that
reversing the sign of any coordinate does not change the layer
of the node.
 Similar to the case of the two-dimensional lattice,
the value of $v_i$ for any node in layer $\ell$
in a $d$-dimensional lattice, denoted by
$v_\ell^{(d)}$, is given by
\begin{equation}
  v_\ell^{(d)} = \left[ T_{\ell_{\max}}^{(d)}(\epsilon) \right]^{-1} \, \epsilon^\ell, \qquad
(0\le \ell\le \ell_{\max}),
\label{eq:vl spatial}
\end{equation}
where
\begin{equation}
T_{\ell_{\max}}^{(d)}(z)
= 1 + \sum_{\ell=1}^{\ell_{\max}} N_{\ell}^{(d)}z^{\ell}
\label{eq:T_lmax^{(d)}(z)}
\end{equation}
From \EQ\eqref{eq:vl spatial}, we obtain
\begin{equation}
\sigma = \frac{\sqrt{T_{\ell_{\max}}^{(d)}(\epsilon^2)}}
{T_{\ell_{\max}}^{(d)}(\epsilon)}.
\label{eq:sigma vs T_lmax^{(d)}}
\end{equation}

Note that $\lim_{\epsilon \to 0} \sigma = 1$ and
$\lim_{\epsilon \to 1} \sigma = N^{-1/2}$. 
In the limit $\ell_{\max}\to\infty$, \EQ\eqref{eq:T_lmax^{(d)}(z)} becomes
\begin{eqnarray}
\lim_{\ell_{\max}\to\infty}T_{\ell_{\max}}^{(d)}(z) 
&=& 1 + \sum_{d^{\prime}=1}^{\infty}
\frac{d!}{d^{\prime}!(d-d^{\prime})!} 2^{d^{\prime}}
\sum_{\ell=d^{\prime}}^{\infty}\frac{(\ell-1)!}
{(d^{\prime}-1)!(\ell-d^{\prime})!}z^{\ell}\nonumber\\
&=& 1+\sum_{d^{\prime}=1}^{\infty}
\frac{d!}{d^{\prime}!(d-d^{\prime})!} 2^{d^{\prime}}
\left(\frac{z}{1-z}\right)^{d^{\prime}}\nonumber\\
&=& \left(\frac{1+z}{1-z}\right)^d.
\label{eq:T_lmax^{(d)}(z) lmax=infty}
\end{eqnarray}
Substituting \EQ\eqref{eq:T_lmax^{(d)}(z) lmax=infty} into
\EQ\eqref{eq:sigma vs T_lmax^{(d)}} yields
\begin{equation}
\lim_{\ell_{\max}\to\infty}\sigma=\left[
\frac{\left(1-\epsilon\right)
\left(1+\epsilon^2\right)}{\left(1+\epsilon\right)^3} \right]^{d/2}.
\end{equation}

\section*{Appendix C: Collective fluctuations in the Cayley tree}

Consider a Cayley tree
with degree $k$ and a specific root node.
We assume that the maximum distance
from the root node is equal to $\ell_{\max}$.
The edges descending from the
root node and those approaching the root node
are assigned weight 1 and $\epsilon$, respectively.
The exact value of 
$v_i$ in layer $\ell$, denoted by $v_{\ell}$ without confusion,
is obtained via
\begin{equation}
\left[1+\left(k-1\right)\epsilon\right]
v_{\ell}=\epsilon v_{\ell-1}+\left(k-1\right)v_{\ell+1},\quad (\ell\ge 1).
\label{eq:linear system tree}
\end{equation}
By solving \EQ\eqref{eq:linear system tree},
we obtain
\begin{equation}
  v_\ell = \frac{1 - (\epsilon k)}{1 - (\epsilon k)^{\ell_{\max}+1}}
 \, \epsilon^{\ell}, \qquad
(0\le \ell\le \ell_{\max}).
\label{eq:vl tree}
\end{equation}
From \EQ\eqref{eq:vl tree}, we obtain
\begin{equation}
\sigma = \frac{1 - (\epsilon k)}{1 - (\epsilon k)^{\ell_{\max}+1}} \,
\sqrt{
\frac{1 - \left( \epsilon^2 k \right)^{\ell_{\max}+1}}
{1 - \left( \epsilon^2 k \right)}
}.
\end{equation}
Note that $\lim_{\epsilon \to 0} \sigma = 1$ and
$\lim_{\epsilon \to 1} \sigma
= \sqrt{(1 - k)/(1 - k^{\ell_{\max}+1})} = N^{-1/2}$.
The infinite-size limit exists only when
$\epsilon k < 1$, and it is equal to
\begin{equation}
\lim_{\ell_{\max}\to\infty} \sigma =
\frac{1-\epsilon k}{\sqrt{1-\epsilon^2 k}}.
\end{equation}

\section*{Appendix D:
Target patterns in continuous media under spatial block noise}

We show that our results for the coupled oscillator system
in the $d$-dimensional lattice are also valid for that in the continuous
Euclidean space. We assume that Gaussian spatial block
noise is applied. This type of noise has been used
in experiments \cite{mikhailov06}.

We consider the $d$-dimensional nonlinear phase diffusion equation given
by
\begin{equation}
  \partial_t \phi(\bm{r}, t)
  = \omega + \nu \nabla^2 \phi + \mu \left( \nabla \phi \right)^2 + s(\bm{r}),
  \label{eq:target 1}
\end{equation}
where $\bm {r}\in \mathbb{R}^d$ is the spatial coordinate,
$\omega > 0$ is the intrinsic frequency,
$\nu > 0$ is the diffusion constant,
and $\mu > 0$ is the coefficient of the nonlinear term~\cite{kuramoto84}.
The term $s(\bm{r})$ represents the localized heterogeneity,
which is positive near the origin and vanishing otherwise.

The synchronous solution corresponding to the target pattern is written as
$\phi(\bm{r}, t) = \Omega t + \psi(\bm{r})$, where $\Omega$ and $\psi(\bm{r})$ satisfy
\begin{equation}
  \Omega
  = \omega + \nu \nabla^2 \psi + \mu \left( \nabla \psi \right)^2 + s(\bm{r}).
  \label{eq:target 2}
\end{equation}
Let $x(\bm r, t)$ be a small deviation from the target pattern
defined by $x \equiv
\phi - (\Omega t + \psi)$. Linearizing \EQ\eqref{eq:target 1} using
$x(\bm r,t)$, we obtain $\partial _t x(\bm{r}, t) = \mathcal{L} x$,
where the linear operator $\mathcal{L}$ is given by
\begin{equation}
  \mathcal{L} x
  = \nu \nabla^2 x + 2 \mu \left( \nabla \psi \right) \cdot \left( \nabla x \right).
\end{equation}
We define the inner product as
\begin{equation}
  \left[ x_1(\bm{r}), x_2(\bm{r}) \right] = \int d\bm{r} \, x_1(\bm r)
   x_2(\bm r).
\end{equation}
We define the adjoint operator $\mathcal{L}^\dag$ as $[x_1,
\mathcal{L}x_2] = [\mathcal{L}^\dag x_1, x_2]$, \textit{i.e.},
\begin{equation}
  \mathcal{L}^\dag x
  = \nu \nabla^2 x - 2 \mu \nabla \cdot \left( x \nabla \psi \right).
\end{equation}
Note that $\mathcal{L}$ is self-adjoint when $\nabla \psi = 0$.

Because of the translational symmetry in \EQ\eqref{eq:target 1}
 with respect to $\phi$, 
$\mathcal{L}$ has one zero eigenvalue.
Let the right and left eigenfunctions of $\mathcal{L}$ corresponding to the zero
eigenvalue be $u(\bm{r})$ and $v(\bm{r})$, respectively, \textit{i.e.}, $\mathcal{L}
u=0$ and $\mathcal{L}^\dag v =0$. Trivially, $u(\bm{r}) = 1$.
The normalization condition $[v(\bm{r}), u(\bm{r})]=1$ then implies that
$\int d\bm{r} \, v(\bm{r}) = 1$.

Now, we introduce the perturbation to \EQ\eqref{eq:target 1} as follows:
\begin{equation}
  \partial_t \phi(\bm{r}, t)
  = \omega + \nu \nabla^2 \phi + \mu \left( \nabla \phi \right)^2 + s(\bm{r})
  + \sqrt{D} \, \xi(\bm{r}, t),
\end{equation}
where $\xi(\bm{r}, t)$ represents a weak perturbation to the target pattern.
Similarly to \EQ\eqref{eq:mode decomposition},
we decompose $x$ into
\begin{equation}
 x(\bm r,t) = y(t) u(\bm r) + \rho(\bm r,t),
\end{equation}
where $y(t) u(\bm r)$ is the collective mode. The dynamical equation for
$y$ is then obtained as
\begin{equation}
  \dot y = \sqrt{D} \int d\bm{r} \, v(\bm{r}) \xi(\bm{r}, t).
  \label{eq:dTheta/dt target}
\end{equation}

Let us assume that $\xi(\bm{r},t)$ is the Gaussian spatial block noise
characterized by
\begin{equation}
  \xi\left( \bm{r}, t \right) = \xi_{\bm{\ell}}\left( t \right), \qquad
  \bm{r} \in \mathbb{R}^d(\bm{\ell}),
  \label{eq:block noise target}
\end{equation}
%
\begin{equation}
  \left\langle \xi_{\bm{\ell}}\left( t \right) \xi_{\bm{\ell}'}\left( t' \right) \right\rangle
  = \delta_{\bm{\ell}, \bm{\ell}'} C\left( \left| t - t' \right| \right),
\end{equation}
where $\bm{\ell}$ is the vector index for the block
$\mathbb{R}^d(\bm{\ell})$.
Using \EQ\eqref{eq:block noise target},
\EQ\eqref{eq:dTheta/dt target} is transformed into 
\begin{equation}
  \dot y = \sqrt{D} \sum_{\bm{\ell}} \, v_{\bm{\ell}} \, \xi_{\bm{\ell}}(t),
\label{eq:dTheta/dt block target}
\end{equation}
where
\begin{equation}
  v_{\bm{\ell}} = \int_{\mathbb{R}^d(\bm{\ell})} d\bm{r} \, v(\bm{r}).
\end{equation}
From \EQ\eqref{eq:dTheta/dt block target}, we find that the intensity of
the collective fluctuation is given by
\begin{equation}
  \sigma = \sqrt{D\sum_{\bm{\ell}} v_{\bm{\ell}}^2}.
\end{equation}
Note that $v_{\bm{\ell}}$ satisfies the normalization condition as follows:
\begin{equation}
  \sum_{\bm{\ell}} v_{\bm{\ell}}
  = \sum_{\bm{\ell}} \int_{\mathbb{R}^d(\bm{\ell})} d\bm{r} \, v(\bm{r})
  = \int d\bm{r} \, v(\bm{r}) = 1.
\end{equation}

\clearpage

\clearpage

\clearpage

\clearpage

\clearpage

\clearpage

\end{document}